%% file: Paper.tex
\documentstyle[epsfig]{europhys}
\input euromacr

\begin{document}
\shorttitle{F. Mancini: The Mott-Hubbard transition in the 2D Hubbard
model}
\title{The Mott-Hubbard transition\\
and the paramagnetic insulating state\\ in the two-dimensional Hubbard
model}
\author{F. Mancini\thanks{E-mail: mancini@vaxsa.csied.unisa.it}}
\institute{Universit\`a degli Studi di Salerno -- Unit\`a INFM di
Salerno\\ Dipartimento di Scienze Fisiche ``E. R. Caianiello''\\ 84081
Baronissi, Salerno, Italy} \pacs{\Pacs{71.10}{Fd}{Lattice Fermion
Models}\Pacs{71.27}{+a}{Strongly Correlated Electronic
Systems}\Pacs{71.30}{+h}{Metal-Insulator Transition}} \euro{}{}{}{} \Date{17 December 1998}
\maketitle

\begin{abstract}
The Mott-Hubbard transition is studied in the context of the
two-dimensional Hubbard model. Analytical calculations show the
existence of a critical value  $U_{c}$ of the potential strength which
separates a paramagnetic metallic phase from a paramagnetic insulating
phase. Calculations of the density of states and double occupancy show
that the ground state in the insulating phase contains always a small
fraction of empty and doubly occupied sites. The structure of the
ground state is studied by considering the probability amplitude of
intersite hopping. The results indicate that the ground state of the
Mott insulator is characterized by a local antiferromagnetic order; the
electrons keep some mobility, but this mobility must be compatible with
the local ordering. The vanishing of some intersite probability
amplitudes at $U=U_{c}$  puts a constrain on the electron mobility. It
is suggested that such quantities might be taken as the quantities
which control the order in the insulating phase.
\end{abstract}

There are several indications that the two-dimensional Hubbard model
(HM) can describe a metal-insulator transition and that some kind of
order is established in the paramagnetic insulating state. However,
there is no clear picture about the structure of the ground state and
no indication about the existence of an order parameter. In particular,
there is a difficulty to conciliate the existence of a finite value for
the doubly occupancy, which implies mobility of the electrons, and the
existence of some order which would imply a localization of the
electrons.

In this article we study the Hubbard model by means of the composite
operator method (COM) in the two-pole approximation. The main results
can be so summarized: (i) a Mott- Hubbard transition does exist; (ii) a
local antiferromagnetic (AF) order is present in the insulating state;
(iii) a quantity which controls the order in the insulating state is
individuated.

According to the band model approximation several transition oxides
should be metals. In practice one finds both metallic and insulating
states, with a metal-insulator transition induced by varying the
boundary conditions (pressure, temperature, compound composition). Mott
[1] pointed out that for narrow bands the electrons are localized on
the lattice ions and therefore the correlations among them cannot be
neglected. A model to describe these correlations was proposed by
Hubbard [2]. In a standard notation his Hamiltonian is given by
\begin{equation}
H=\sum_{ij}(t_{ij} - \mu\delta_{ij}) c^\dagger(i) c(j) +U\sum_{i}
n_{\uparrow}(i) n_{\downarrow}(i)
\end{equation}       
where $c(i)$ and $c^\dagger(j)$  are annihilation and creation
operators of electrons at site $i$,  in the spinor notation;  $t_{ij}$
describes hopping between different sites and it is usually taken as
$t_{ij}=-4t\alpha_{ij}$, where $\alpha_{ij}$  is the projection
operator on the first neighbor sites; the  $U$-term is the Coulomb
repulsive interaction between two electrons at the same site with
$n_{\sigma}(i) = c^\dagger_{\sigma}(i) c_{\sigma}(i)$; $\mu$ is the
chemical potential.

The magnitudes of the on-site Coulomb energy $U$ and the
one-electron band width $W=8t$ control the properties of the
system. In this competition between the kinetic and the potential
energy the most difficult part of the model resides and exact
solutions do not exist, except in some limiting cases. In
particular, an adequate description of the  ground state and
elementary excitations is still missing. In the case of one
dimension an exact solution is available [3] which shows that
there is no Mott-Hubbard transition: a gap in the density of
states is present for any value of U. In higher dimensions there
are several results that indicate the existence of  a Mott-Hubbard
transition, in the sense that at half filling there is a critical
value of the Coulomb potential $U_{c}$  which separates the
metallic phase from the insulating phase; but no rigorous results.
In Hubbard I approximation [2] and in the work by Roth [4] no
transition is observed [5]. In Hubbard III approximation [6] an
opening of the gap is observed for $U_{c}=W\sqrt{3/2}$. By using
the Gutzwiller variational method [7] Brinkman and Rice [8] find
$U_{c}=8|\overline\epsilon |\approx 1.65 W$, with
$\overline\epsilon$ being the average kinetic energy per electron;
the vanishing of the double occupancy D at this value induced them
to propose the double occupancy as an order parameter to describe
the metal-insulator transition. However, this result is based on
the use of the Gutzwiller approximation, which becomes exact only
for infinite dimensions [9]. For finite dimensions theoretical
[10] and numerical analysis [11] show that the double occupancy
tends to zero only in the limit $U\to \infty$. By using the
dynamical mean-field approach, or  $d\to\infty$ limit, Georges et.
al. [12] find that at some critical $U$ a gap opens abruptly in
the density of states, due to the disappearance of a Kondo-like
peak. A recent calculation [13] of the HM in infinite dimensions
shows a continuous Mott-Hubbard transition, with a gap opening at
$U_{c}\approx W$. The same qualitative result has been found [14]
by using Quantum Monte Carlo (QMC) simulation; working at the high
$T=0.33\, t$  the authors observe a transition with a gap opening
continuously at $U_{c}\approx W/2$.

In conclusion, while there are several results indicating the
existence of a Mott-Hubbard transition in the 2D Hubbard model,
there is no unified picture; the mechanisms that lead to the
transition are different; the value of the critical interaction
strength varies from $U_{c}\approx 0.5W$ [14] up to $U_{c}\approx
1.65W$  [8]. A description of the structure of the ground state in
the paramagnetic insulating state is also lacking.

In the framework of the COM [15,5] the Hubbard model has been
solved in the two pole approximation [16], where the operatorial
basis is described by the doublet Heisenberg operator
\begin{equation}
\psi (i) = \left(\begin{array}{c} \xi(i)\\ \eta(i)
\end{array} \right)
\end{equation}       
and finite life-time effects are neglected. The fields $\xi
(i)=[1-n(i)] c(i)$ and $\eta(i)=n(i) c(i)$, with $n(i)= c^\dagger (i)
c(i)$, are the Hubbard operators [2]. In this framework the
single-particle propagator is given by
\begin{equation}
F.T. \langle R[\psi(i)\psi^\dagger(j)]\rangle
=\sum^{2}_{i=1}\frac{\sigma^{(i)}({\bf k})}{\omega-E_i({\bf k})+i\eta}
\end{equation}       
where  $F.T.$ means Fourier transform. The expressions of the
spectral functions  $\sigma^{(i)}(\mathbf{\mathrm{k}})$ and energy
spectra $E_{i}(\mathbf{\mathrm{k}})$ have been reported in
previous works [15]. These functions are calculated in a fully
self-consistent treatment, where attention is paid to the
conservation of relevant symmetries [15,5,17]. Differently from
other approaches, one does not need to recur to different schemes
in order to describe the weak- and strong-coupling regimes. {\it
Both the limits $U\to 0$ and $U\to\infty$  are recovered by Eq.
(3).}

The result (3) has been derived by assuming a paramagnetic phase.
It is an open question if this is the true ground state at half filling
and zero temperature. Results of numerical simulation seem to indicate
that the paramagnetic phase is unstable against a long range antiferromagnetic
order. However, numerical analysis is severely restricted in cluster size, and it
is very hard to conclude that the true solution has an infinite range AF order. As
it will be shown later, the calculation of the probability amplitudes for electron
transfers shows that in the paramagnetic insulating state a local AF order is established,
with a correlation length of the order of few hundred time the lattice constant.

At first, we observe that the Mott-Hubbard transition can be studied by
looking at the chemical potential, which is the quantity which mostly
controls the single-particle properties. Let us define
\begin{equation}
-\mu_{1} = \left({\partial\mu\over \partial n}\right) _{n=1} = {1\over
\kappa (1)}
\end{equation}       
where  $\kappa (n) = (\partial n/\partial\mu)/n^{2}$ is the
compressibility. Analytical calculations show that at zero temperature
there is a critical value of the interaction, fixed by the equation
\begin{equation}
U_{c} = 8t\sqrt{4p-1}
\end{equation}       
such that for $U>U_{c}$  the quantity $\mu_{1}$  diverges. The
parameter  $p$  describes a bandwidth renormalization and is defined by
\begin{equation}
p\equiv {1\over 4} \langle n^\alpha_{\mu}(i) n_{\mu} (i)\rangle -
\langle [ c_{\uparrow}(i) c_{\downarrow}(i)]^\alpha
c_{\downarrow}^\dagger (i) c^\dagger_{\uparrow}(i) \rangle
\end{equation}       
where $n_{\mu}(i)=c^\dagger (i)\sigma_{\mu} c(i)$  is the charge
($\mu=0$) and spin ($\mu=1,2,3$) density operator. We use the notation
$A^\alpha(i)=\sum_{j} \alpha_{ij} A(j)$  to indicate the operator $A$
on the first neighbor site of $i$. The quantities $p$ and $\mu_{1}$ are
functions of the external parameters $n$, $T$, $U$ and are
self-consistently calculated. Numerical solution of the self-consistent
equation (5) shows $U_{c} \approx 1.68W$. In Fig. 1a $\mu_{1}$  is
plotted versus $U/t$ for $k_{B} T/t=0,0.3,1$. At finite temperature
$-\mu_{1}$  increases by increasing U and tends to $\infty$  in the
limit $U\to \infty$. At zero temperature $\mu_{1}$  exhibits a
discontinuity at $U=U_{c}$.When the intensity of the local interaction
exceeds the critical value $U_{c}$, the chemical potential exhibits a
discontinuity at half filling, showing the opening of a gap in the
density of states and therefore a phase transition from the metallic to
the insulating phase.
\begin{figure}
\begin{center}
\parbox[b][8cm][s]{6.5cm}{\epsfig{file=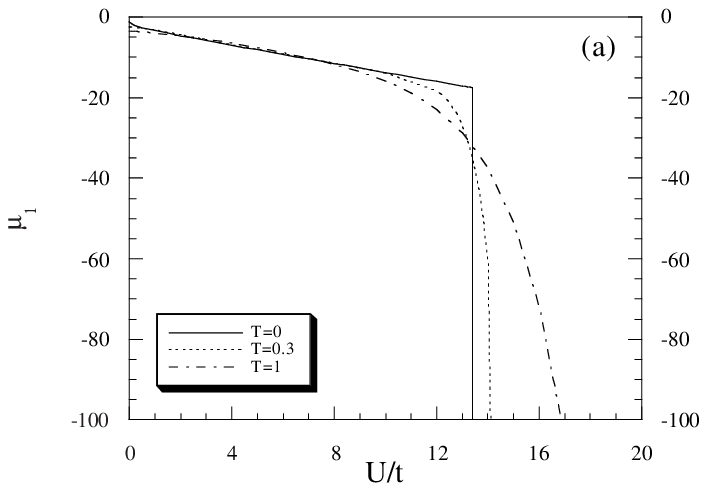,width=6cm,clip=}\vfill
\epsfig{file=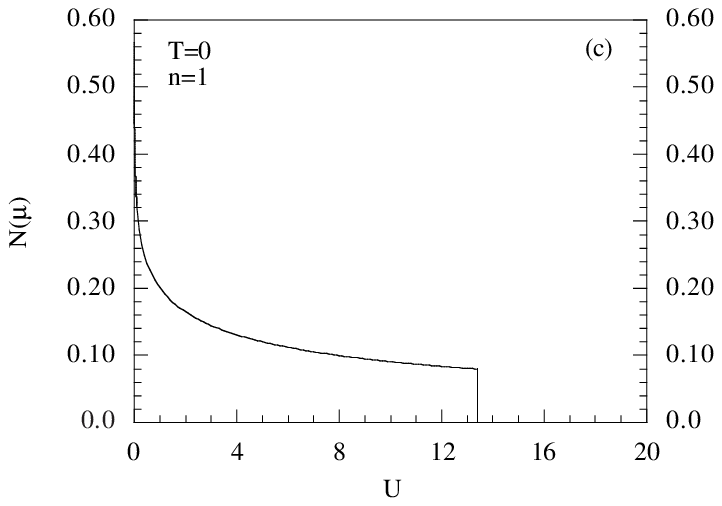,width=6cm,clip= }}\hfill
\epsfig{file=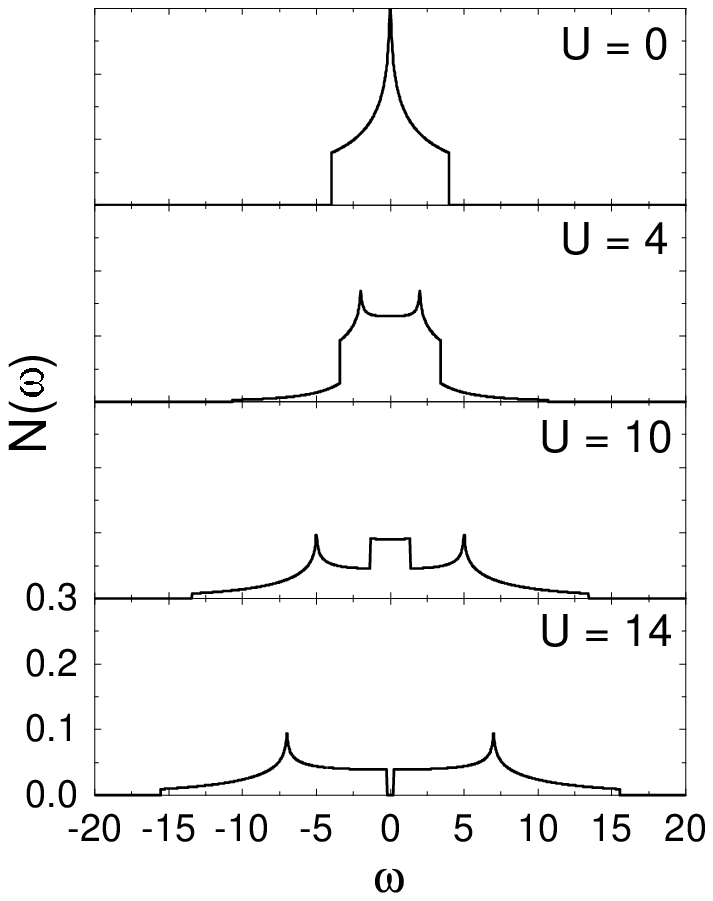,height=7cm,clip= }
\end{center}
\caption{(a) The parameter $\mu_{1}=-(\partial\mu/\partial n)_{n=1}$ is
plotted as a function of the potential strength $U/t$  for various
values of the reduced temperature $k_{B}T/t$; (b) The electronic
density of states at the Fermi value is given against the potential
strength; (c) The electronic density of states  is plotted as a
function of the energy for half filling and different values of U.
$U/t$.}
\end{figure}
Calculations show that the density of states (DOS) is made up by two bands: lower and upper
band. When   $U<U_{c}$ the two band overlap: metallic phase. The region
of overlapping is given by $\Delta\omega = 16 tp-\sqrt{U^{2}+64t^{2}
(2p-1)^{2}}$. When  $U>U_{c}$ the two band do not overlap: insulating
phase. In Fig. 1b the electronic DOS is reported for different values
of $U$. We see that when $U$ increases the central peak opens in two
peaks; some of the central weight is transferred to the two peaks,
which correspond to the elementary excitations, described by the fields
$\xi$ and $\eta$. When  $U$ reaches the critical value $U_{c}\approx
1.68 W$ the central peak vanishes abruptly; a gap appears and the
electronic density of states splits into two separate bands. This is
seen in Fig. 1c, where the DOS calculated at the Fermi level is plotted
versus $U$. We find that the gap develops continuously, following the
law $\Delta \approx 1.5W (U/U_{c}-1)$.

A more detailed study of the density of states can be obtained by
considering the contributions of the different channels. Calculations
show that both the fields $\xi$ and $\eta$ contribute to the two bands.
Only in the limit $U\to\infty$ the two operators do not interact and
separately contribute to the two bands. Although, the lower band is
essentially made up by the contribution of ``$\xi$-electron", there is
always a contribution coming from the ``$\eta$-electron". The viceversa
is true for the upper band. Particularly, the cross contribution plays
an important role in the region around the Fermi value, where
$N_{\xi\eta} (\mu) \approx N_{\xi\xi}(\mu) \approx N_{\eta\eta} (\mu)$
[for $U>0$].

This result shows that in the insulating phase the ground state has a
structure different from the simple one where all sites are singly
occupied; the competition between the itinerant and local terms leads
to a ground state characterized by a small fraction of empty and doubly
occupied sites. Some questions arise: (1) what is the structure of the ground state? and in particular
there exists any order?;
(2) if an ordered state is established, why this order is not destroyed by the mobility of the
electrons;
(3) can we individuate an order parameter describing the transition at $U=U_{c}$?
\begin{figure}
\begin{center}
\epsfig{file=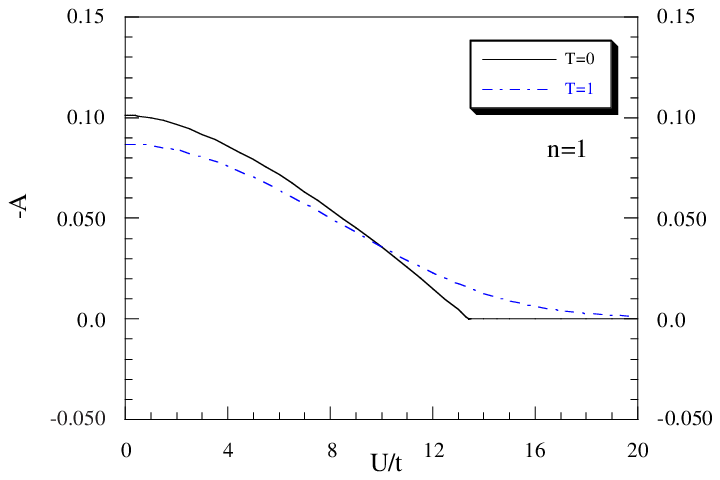,height=4.5cm,clip=} \hfill
\epsfig{file=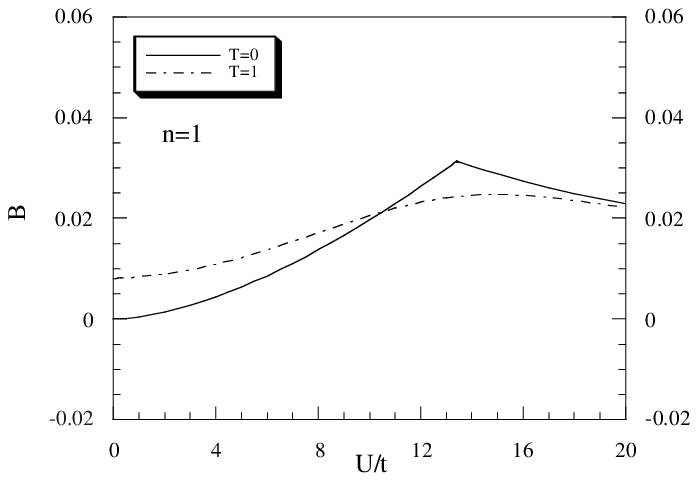,height=4.5cm,clip=}
\end{center}
\caption{(left) The first nearest neighboring hopping amplitude
$A^\alpha= \langle\xi^\alpha (i) \xi^\dagger (i)\rangle =
\langle\eta^\alpha (i) \eta^\dagger (i)\rangle$   is reported as a
function of  $U/t$ for zero temperature and for $k_{B} T/t=1$. (right)
The first nearest neighboring hopping amplitude
$B^\alpha=\langle\eta^\alpha (i) \xi^\dagger (i)\rangle =
\langle\xi^\alpha (i) \eta^\dagger (i)\rangle$ is reported as a
function of $U/t$  for zero temperature and for $k_{B} T/t=1$.}
\end{figure}
An important quantity for the comprehension of the properties of
the system is the double occupancy $D=\langle n_{\uparrow}
n_{\downarrow}\rangle$  which gives the average number of sites
occupied by two electrons. Analytical calculations show that at
zero temperature the double occupancy, as a function of $U$,
exhibits  a drastic change when the critical value is crossed,
however remains finite for $U>U_{c}$ and tends to zero only in the
limit of infinite $U$ as $\lim_{U\to\infty} D = J/ 8U$ where
$J=4t^{2}/U$ is the AF exchange constant. In the case of one
dimension our analytical results give $\lim_{U\to\infty} D =
3t^2/U^2$ which is very close to the Bethe Ansatz result
$\lim_{U\to\infty} D = 4\ln 2t^2/ U^2$ [18]. Double occupite sites
are used by the system in order to lower its energy. As a matter
of fact this is precisely the origin of the effective spin-spin
interaction in the $t-J$ model [19]. To better understanding the
structure of the ground state, we have to study the matrix element
$\langle c_{\sigma} (j) c^\dagger_{\sigma}(i)\rangle$. This
quantity represents the probability amplitude that an electron of
spin $\sigma$ is created at site $i$ and an electron of spin
$\sigma$ is destroyed at site $j$.  However, this quantity gives
only a limited information about the occupation of the sites $i$
and $j$; there are four possible ways to realize the transition
$j(\sigma)\to i(\sigma)$, and the quantity $\langle c_{\sigma}(j)
c_{\sigma}^\dagger(i)\rangle$  cannot distinguish among them. By
means of the decomposition $c_{\sigma }(i) = \xi_{\sigma} (i) +
\eta_{\sigma}(i)$, the probability amplitude is written as the sum
of four contributions $\langle c_{\sigma}(j)
c_{\sigma}^\dagger(i)\rangle = \langle \xi_{\sigma} (j)
\xi^\dagger_{\sigma} (i)\rangle + \langle\xi_{\sigma} (j)
\eta^\dagger_{\sigma}(i)\rangle +\langle\eta_{\sigma} (j)
\xi^\dagger_{\sigma}(i)\rangle + \langle\eta_{\sigma} (j)
\eta^\dagger_{\sigma}(i)\rangle$ which correspond to the following
transitions:
\begin{center}
$\begin{array}{cccccc} \langle \xi_{\sigma }(j) \xi^\dagger_{\sigma}
(i) \rangle :\qquad & \framebox{0}\ &\  \framebox{$\sigma$}\ & \ \to\ &
\framebox{$\sigma$}\ &  \framebox{0}\\ & \raisebox{6pt}{\it i} &
\raisebox{6pt}{\it j} && \raisebox{6pt}{\it i}  & \raisebox{6pt}{\it j}
\\ \langle \xi_{\sigma }(j) \eta^\dagger_{\sigma} (i) \rangle :\qquad &
\framebox{$-\sigma$}\ &\ \framebox{$\sigma$}\ &\ \to\ &
\framebox{$\sigma-\sigma$}\ & \framebox{0}\\ & \raisebox{6pt}{\it i} &
\raisebox{6pt}{\it j} && \raisebox{6pt}{\it i}  & \raisebox{6pt}{\it j}
\\ \langle \eta_{\sigma }(j) \xi^\dagger_{\sigma} (i) \rangle :\qquad &
\framebox{0}\ &\ \framebox{$\sigma-\sigma$}\ &\ \to\ &
\framebox{$\sigma$}\ & \framebox{$-\sigma$}\\ & \raisebox{6pt}{\it i} &
\raisebox{6pt}{\it j} && \raisebox{6pt}{\it i}  & \raisebox{6pt}{\it j}
\\ \langle \eta_{\sigma}(j) \eta^\dagger_{\sigma} (i) \rangle :\qquad &
\framebox{$-\sigma$}\ &\ \framebox{$\sigma-\sigma$}\ &\ \to\ &
\framebox{$\sigma-\sigma$}\ &  \framebox{$-\sigma$}\\ &
\raisebox{6pt}{\it i} & \raisebox{6pt}{\it j} && \raisebox{6pt}{\it i}
& \raisebox{6pt}{\it j}
\end{array}$
\end{center}
A study of the probability amplitudes $\langle\psi (j) \psi^\dagger
(i)\rangle$  will give detailed information about the structure of the
ground state. In Fig. 2a the amplitude  $A= \langle\xi^\alpha (i)
\xi^\dagger (i)\rangle = \langle\eta^\alpha (i) \eta^\dagger
(i)\rangle$ is reported as a function of $U/t$ for two different
temperatures $k_{B}T/t = 0,1$. We see that in the case of zero
temperature this quantity vanishes for $U>U_{c}$. This can be easily
seen also by analytical methods. The quantity  $B=\langle\eta^\alpha
(i) \xi^\dagger (i)\rangle = \langle\xi^\alpha (i) \eta^\dagger
(i)\rangle$ is reported in Fig. 2b; we see that this probability
amplitude does not vanish above $U_{c}$.

Owing to this contribution, we have that for $U>U_{c}$  the hopping of
electrons from site $i$ to the nearest neighbor is not forbidden,
although restricted by the fact that $A=0$. The hopping amplitudes have
been studied up to the third nearest neighbors, but the analysis is
easily extended to any site, by symmetry considerations. The scheme
that emerges from analytical and numerical calculations can be
summarized in the following table.
\begin{center}
\epsfig{file=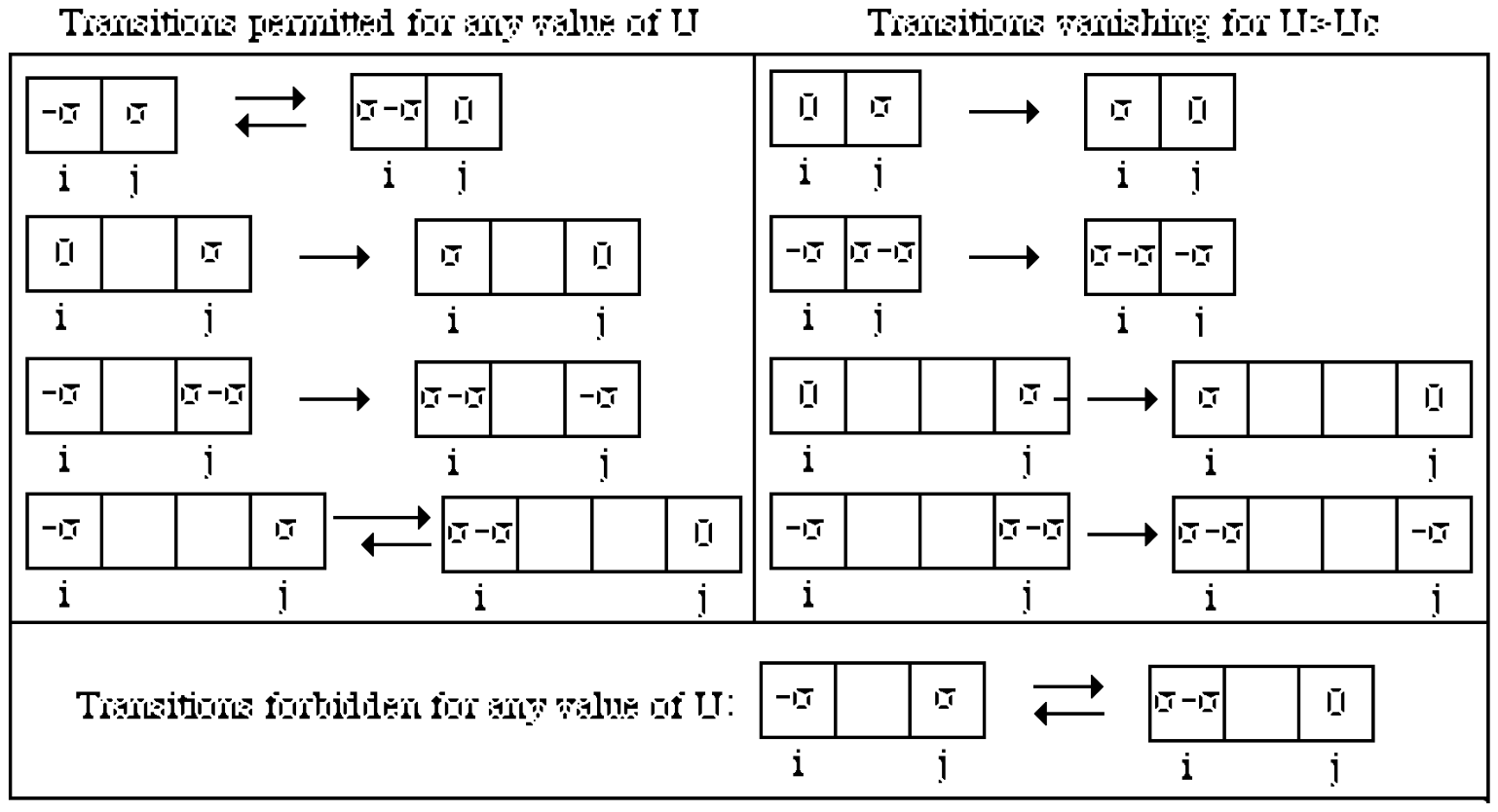,width=10cm,clip=}
\end{center}
Putting all these results together, for $U>U_{c}$  the situation can be
so summarized:
\begin{enumerate}
\item an electron  $\sigma$ which singly occupies a site
(a) can hop on first, third,.....neighboring sites if and only if these sites are already occupied
by an electron $-\sigma$;
(b)  can hop on second, fourth,.....neighboring sites if and only if these sites are empty;
\item an electron $\sigma$  which doubly occupies a site
(a) can hop on first, third,.....neighboring sites if and only if these sites are empty;
(b)  can hop on second, fourth,.....neighboring sites if and only if these sites are already
occupied by an electron $-\sigma$.
\end{enumerate}
The picture that emerges by these results is that the paramagnetic
ground state in the insulating phase is characterized by a finite-range
antiferromagnetic order. Due to the fact that there are empty and
doubly occupied sites, the electrons have some mobility, but there are
strong constrains on this mobility, such that the local
antiferromagnetic order is not destroyed. This result is consistent
with the fact that there is a competition between the itinerant and
localizing energy terms. A study of the kinetic and potential energies
as functions of U shows that for any $t\neq 0$  there is always some
contribution which comes from the kinetic energy which allows the
hopping among sites. Only in the limit of infinite U, the double
occupancy and all transition amplitudes go to zero.

In conclusion, the two-dimensional single-band Hubbard model at
half filling and zero temperature has been studied by means of the
composite operator method. Analytical calculations show the
existence of a critical value $U_{c}$ which separates the metallic
and insulating phases. As soon as U increases from zero, a
depletion appears in the density of states; some weight of the
central region is transferred to the lower and upper Hubbard
bands. For larger values of U, DOS develops three separated
structures: part of the weight remains in the center around the
Fermi value and discontinuously disappears at $U=U_c$. Similar
results, although based on different mechanism, have been
previously obtained in Ref. [20] for the case of
infinite-dimensional Hubbard model, in Ref. [21] by means of
standard perturbation expansions, in Ref. [22] by Monte Carlo
simulations. For  $U>U$ a gap opens and the density of states
splits into two separated structures.  Our calculations show that
even for $U\gg U_{c}$, where the lower and upper bands are well
separated, the two contributions coming from $\xi$ and $\eta$ do
not separate. The ground state in the insulating phase contains
always a small fraction of empty and doubly occupied sites.

This result is confirmed by the study of the matrix element $\langle
c_{\sigma} (j) c^\dagger _{\sigma}(i)\rangle$, which gives the
probability amplitude of hopping from the site $j$ to the site $i$.
When $j$ is an odd nearest neighboring site of $i$, this
quantity is not zero for $U>U_{c}$  and vanishes only for infinite U.
However, when we split $c=\xi+\eta$  and analyze $\langle c_{\sigma}
(j_{odd}) c^\dagger_{\sigma}(i)\rangle$ in components, we find that for
$U>U_{c}$  only the matrix elements  $\langle \xi_{\sigma} (j_{odd})
\eta^\dagger_{\sigma}(i)\rangle$ and $\langle \eta_{\sigma} (j_{odd})
\xi^\dagger_{\sigma}(i)\rangle$ survive. The probability amplitudes
$\langle \xi_{\sigma} (j_{odd}) \xi^\dagger_{\sigma}(i)\rangle$ and
$\langle \eta_{\sigma} (j_{odd}) \eta^\dagger_{\sigma}(i)\rangle$
vanish at $U=U_{c}$   and remain zero for all $U>U_{c}$. On the other
hand, the matrix element $\langle c_{\sigma} (j_{even})
c^\dagger_{\sigma}(i)\rangle$    is always zero for any value of U; the
two contributions $\langle \xi_{\sigma} (j_{even})
\xi^\dagger_{\sigma}(i)\rangle$ and $\langle \eta_{\sigma} (j_{even})
\eta^\dagger_{\sigma}(i)\rangle$ compensating each other.

Summarizing, our calculations suggest that the ground state of the Mott
insulator has the following characteristics:
(1)  a small fraction of sites are empty or doubly occupied; the number of these sites depend on
the value of U/t and tends to zero only in the limit $U\to \infty$;
(2)  a local antiferromagnetic order is established;
(3)  the electrons keep some mobility, but this mobility must be compatible with the local AF
order;
(4)  the matrix elements $\langle \xi_{\sigma} (j_{odd}) \xi^\dagger_{\sigma}(i)\rangle$
and $\langle \eta_{\sigma} (j_{odd}) \eta^\dagger_{\sigma}(i)\rangle$
might be considered as the quantities which control the order in the
insulating phase.

\stars

The author wishes to thank Doctors Adolfo Avella and  Dario Villani for
valuable discussions. It is gratefully acknowledged an enlightening
discussion with Professor Peter Fulde, that partly motivated the
writing of this article.

\end{document}

%% file: euromacr.tex
\def\stars{\bigskip\centerline{***}\medskip}

\newif\ifboo \boofalse


%% file: Paper.bbl
\begin{thebibliography}{99}
\bibitem{} We wish to thank the referee for this remark.
\end{thebibliography}
